\documentclass[11pt,psfig,emulateapj]{emulateapj}
\usepackage{lscape}
\usepackage{graphics}

\newcommand{\etal}{et al.}

\input epsf 

\newcommand{\LCDM}{{$\Lambda$CDM }}
\def\be{\begin{equation}}\def\bea{\begin{eqnarray}}\def\beaa{\begin{eqnarray*}}
  \def\ee{\end{equation}}  \def\eea{\end{eqnarray}}  \def\eeaa{\end{eqnarray*}}
\def\MPC{\rm{Mpc}}
\def\LCDMl{\rm{$\Lambda$CDM-l}}
\def\LCDMs{\rm{$\Lambda$CDM-s}}
\slugcomment{Not to appear in Nonlearned J., 45.}

\shorttitle{New Halo Finding Method for N-body Simulations}
\shortauthors{Kim and Park}

\begin{document}
\title{A New Halo Finding Method for N-body Simulations}
\author{Juhan Kim\altaffilmark{1,2} and Changbom Park\altaffilmark{3,4}}

\altaffiltext{1}{Korea Astronomy Observatory 61-1, Whaam-Dong, 
Youseong-Gu, Daejon, 505-548, Korea}
\altaffiltext{2}{kjhan@kao.re.kr}
\altaffiltext{3}{Korea Institute for Advanced Study 207-43, 
Cheongnyangni 2-Dong, Dongdaemun-Gu, Seoul, 130-722, Korea}
\altaffiltext{4}{cbp@kias.re.kr}



\def\hfind{PSB }
\def\PS{\rm PS}
\def\LCDM{${\rm \Lambda CDM}$~}
\def\LCDMl{${\rm \Lambda CDM}$-$l$}
\def\LCDMs{${\rm \Lambda CDM}$-$s$}
\begin{abstract}
We have developed a new halo finding method,
Physically Self-Bound (PSB) group finding algorithm,
which can efficiently identify halos located even at crowded regions.
This method combines two physical criteria such as the tidal radius
of a halo and the total energy of each particle to find member particles.
No subtle dependence of halo mass functions 
on various parameters of the PSB method has been found.
Two hierarchical meshes are used to increase 
the speed and the power of halo identification
in the parallel computing environments.
First, a coarse mesh with cell size equal to the mean particle separation
$l_{\rm mean}$
is used to obtain the density field over the whole simulation box.
Mesh cells having density contrast higher than
a local cutoff threshold $\delta_{\rm LOC}$ are extracted and
linked together for those adjacent to each other.
This produces local-cell groups.
We analyze the group of 
particles located at each local-cell group region separately.
This treatment makes the halo finding method easily implemented on 
the parallel computing environments 
since each computational rank takes the halo identification job 
in each particle group independently.
Second, a finer mesh is used to obtain density field 
within each local-cell group and to identify halos.
We set the cell size of the refined mesh to be twice 
the gravitational force softening length $\epsilon$.
The density peaks in the fine mesh are the halo candidates.
Based on the fine mesh, we split
particles into many groups located in the density shells 
with different density levels.
If a density shell contains only one density peak,
its particles are assigned to the density peak.
But in the case of a density shell surrounding at least two density peaks,
we use both the tidal radii of halo candidates enclosed by the shell and
the total energy criterion to find physically bound particles with respect to
each halo.

We have tested the \hfind using a binary halo model 
against other popular halo finding methods, 
such as the Friend-of-Friend (FoF), 
Spherical Overdensity (SO), DENMAX, and HOP.
Similar to DENMAX and HOP,
the \hfind method can efficiently identify 
small halos embedded in a large halo,
while the FoF and the SO do not resolve such small halos.
We apply our new halo finding method to a 1-Giga particle simulation
of the $\Lambda$CDM model
and compare the resulting mass function with those of previous studies.
The abundance of physically self-bound halos
is larger at the low mass scale and smaller at the high mass scale
than proposed by the Jenkins et al. (2001) 
who used the FoF and SO methods.

\end{abstract}

\section{Introduction}
Cosmological N-body simulations (Park 1990; Gelb \& Bertschinger 1994;
Park 1997;
Evrard et al. 2002; Dubinksi et al. 2003; Bode \& Ostriker 2003, among others) 
have been used to test cosmological models
in various fields of interests. 
To compare simulation results with observed galaxies or 
other visual cosmic objects,
one has to extract virialized dark matter halos from the distribution 
of simulation particles.

An easy way to identify virialized halos 
from particle distribution is 
to link particles with distances less than 
$l_{\rm FoF}$ (Friend-of-Friend, Audit et al. 1998; Davis et al. 1985).
The value of the linking length $l_{\rm FoF}$
is usually set to $0.2 \times l_{\rm mean}$ corresponding to
the virialization overdensity $\rho/\rho_b=178$ (Porciani et al. 2002),
where $\rho_b$ is the mean background density.
The chain of linked particles forms a group of particles and 
the particle group is considered as a virialized halo.
But the FoF cannot identify small halos embedded in 
larger high-density regions.
This effect is similar to the ``overmerging'' problem occurred 
in the poor-resolution simulations (Moore et al. 1996; Klypin et al. 1999).
The main drawback of the FoF is 
the ``overlinking'' (Gelb \& Bertschinger 1994) 
since, in a binary or multiple halo system, member halos are often linked
by bridging particles.
Then a dumbbell-like halo is identified 
and, consequently, the halo quantities, such as the center of mass,
the shape, the rotation velocity, etc., are blended.
The Spherical Overdensity (SO, Lacey \& Cole 1994; Warren et al. 1992)
uses the mean density for virialized halos.
The SO searches for density peaks and puts spheres around them by increasing
the radius of the sphere until the internal mean density
satisfies the virialization criterion ($\rho/\rho_b=178$).
Particles inside the sphere are grouped as members of the spherical halo.
However, small halos embedded in virialized regions
can not be properly resolved.

More improved halo finding methods have been proposed in the last decade.
The particle sliding scheme in the density field
is adopted by the DENMAX (Gelb \& Bertschinger 1994)
or its variant SKID (Weinberg et al. 1997; Jang-Condell \& Hernquist 2001),
and the HOP (Eisenstein \& Hut 1998).
The DENMAX uses rectangular density grid cells to slide particles to the nearby
densest grid cell.
Then it scoops up particles that are stacked at local density maxima
and checks the total energy of each particle 
to discard unbound particles.
The SKID method, an improved version of the DENMAX, 
uses a variable smoothing length and the density gradient in a
coordinate-free density field.
The HOP calculates the density field in the way similar to the SKID.
However, it uses a different type of particle sliding.
The HOP method searches for the maximum density 
among a particle's nearest neighbors.
Particles are slid into the nearby densest particle.
The HOP groups particles in local density maxima as virialized halos
similar to the DENMAX.

But above halo finding methods are mainly based only on 
the density field
in determining the member particles of halos.
Namely, used are the density-related quantities 
such as the linking length (FoF),
the overdensity (SO), and 
the sliding on a local density field (DENMAX, SKID, and HOP). 
As a result, the distribution of particles in a halo depends on 
the density geometry.
For example, the FoF is likely to produce dumbbell-like halos and
the SO finds halos with spherical boundaries.
Both methods cannot resolve any substructure inside the virialized region.
Using the particle sliding scheme
seems more proper than the overdensity criteria 
($\rho/\rho_b$ and $l_{\rm FoF}$) since it is reasonable to assume that
boundaries of virialized halos may follow 
the 3-dimensional density valley when they are next to other halos.
But, in some cases,
the density gradient field does not properly describe halo boundaries
when there is no density valley.
For example, consider a small halo is located in a large halo.
Around the small halo, there may exist a density valley 
between the two density peaks of halos.
However, there is no density valley 
in the outskirts of the large halo region beyond the small halo.
So the density valley does not form a closed surface.
Consequently, particles that are members of the large halo may slide 
to the small halo along the path 
without encountering any density valley.

In this paper, we present a new halo finding method
based on the density map.
However, our method also takes into account 
the tidal radius that forms a closed surface around the halo
and the total energy to select physically bound particles only.
We describe how to divide simulation particles into local particle groups 
in \S\ref{ilgg}.
For individual particle groups, we present 
the way to identify virialized and stable halos in the tidal field
in \S\ref{psh}.
The values of the group finding parameters are given in 
 \S\ref{vdeltal},
and we test the sensitivity of the \hfind to parameters in \S\ref{parasec}.
In \S\ref{hfcomp}, we compare our halo finding method
with other popular methods.
In \S\ref{hfresult}, we apply our method to the 1-Giga particle simulations.
Conclusions follow in \S\ref{hfcon}.

\section{Local Particle Groups on a Coarse Mesh}
\label{ilgg}
The \hfind is based on the two hierarchical meshes for particle-density field.
A low-resolution density field is used to find the ``local particle groups'',
and for each local particle group
we use a high-resolution density field to find density peaks and
to divide particles into multiple ``particle sets''.
The division of whole simulation particles into many local particle groups
allows us to reduce computational costs in halo findings.

The \hfind uses all simulation particles
to build the density field
on a coarse mesh with cell size equal to the mean inter-particle separation
$l_{\rm mean}$.
To assign densities to the mesh cells 
we use the $W_4$ (Monaghan \& Lattanzio 1985) 
\begin{equation}
W_4(r,h) = {1 \over \pi } \cases{(1- {3\over 2} x + {3\over 4} x^3)
             &if $0\le x < 1$,\cr
           {1\over 4} (2-x)^3  &if $1\le x < 2$,\cr
             0  &otherwise,\cr}
\end{equation}
where $r$ is the distance of a particle to a grid point,
$h$ is the smoothing length, and $x\equiv r/h$.
We set $h$ equal to $l_{\rm mean}$ in the coarse mesh.
We attach particles to mesh cells as the linked list (Hockney \& Eastwood 1981).
We extract overdense cells 
enclosed by an isodensity surface defined by the density contrast
threshold $\delta_{\rm LOC}$. These cells are grouped and labeled. 
To these grouped cells, we attach adjacent underdense mesh cells
to avoid sharp boundary truncations of halos.
The particles in a local-cell group are found and called 
the local particle group.

\section{Particle Set Hierarchy and Particle Assignment}
\label{psh}

\subsection{Local Density Field on a Fine Mesh}
\label{denf}
For each local particle group, we construct 
a high resolution density field on a fine mesh with cell size and smoothing length 
equal to twice the force softening length $\epsilon$ 
to resolve small but physically meaningful structures.
We search for local density peaks above the density contrast threshold $\delta_p$.
Those peaks are called halo cores.
Around each density peak, 
we find the minimum density level $\delta_c$ with which the isodensity
surface encloses only that peak (see Fig. 1).
Particles in cells surrounded by this isodensity surface
are set to be core members of the density peak.
We constrain that the number of core particles ($N_{core}$) 
should be at least 10 to avoid the detection of spurious peaks 
induced by the Poisson fluctuation.
We assume those core particles as members of the halo.
Other remaining particles are divided into subsets located within
the density shells having $N_{level}$ density levels
between $\delta_{\rm LOC}$ and $\delta_{c,max}$, the highest $\delta_c$.
Every density shell encloses at least two halo cores.
We call the group of particles in a density shell a particle set (PS).
Now we have to assign each of these particles 
to one of the halo cores.

\begin{figure}[tbp]
\begin{minipage}[t]{7cm}
\leavevmode
\epsfxsize=8cm
\epsfysize=9cm
\epsffile{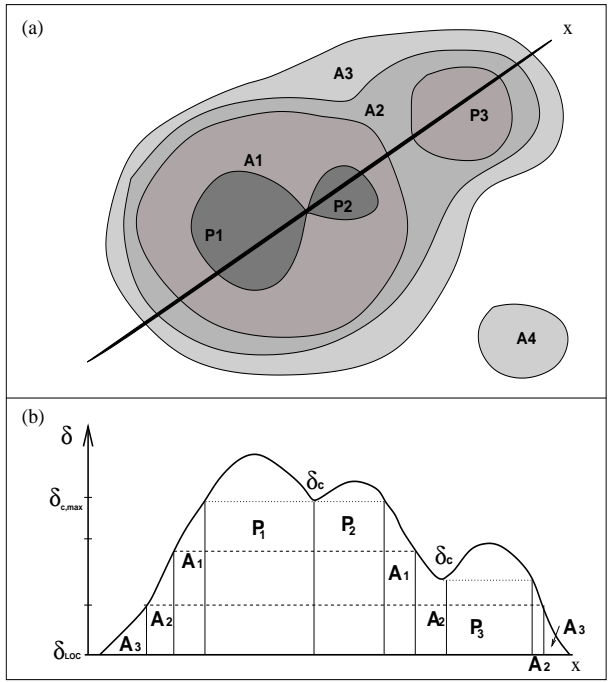}
\end{minipage}
\caption
{
Schematic illustration of a density map in our halo finding algorithm.
In panel (a),
The regions {\bf P1, P2, P3} and {\bf A4} contain only one density peak
with $\delta \ge \delta_p$, and are enclosed by isodensity contour surfaces.
The regions {\bf A3} and {\bf A4} are identified by the density threshold
$\delta_{\rm LOC}$.
In panel (b), we show the density map along the line x in panel (a).
A horizontal dotted line shows the core density contrast $\delta_c$
of each peak.
We use $N_{level}=3$ and show the density level in a horizontal dashed line.
}
\label{denmap}
\end{figure}

We illustrate our halo finding method in Figure \ref{denmap}.
Shown are two regions surrounded by the outermost density contours
with the threshold level $\delta_{\rm LOC}$.
As described in the previous section we call each region
a local-cell group.
The boundaries of the inner-most density region, 
{\bf P1}, {\bf P2}, and {\bf P3}
are found by searching for $\delta_c$'s.
Particles in those regions are the core members
of each density peak.
Now we call each of {\bf P1}, {\bf P2}, and {\bf P3} a halo candidate.
To simplify the illustration in Figure 1., 
we use three density threshold ($N_{level}=3$)
and get three particle sets spatially split as {\bf A1}, {\bf A2}, and {\bf A3}.
But actually, we use $N_{level}=10$.
Particles in the region {\bf A1} 
may be assigned to one of {\bf P1} and {\bf P2}.
Particles in the regions of {\bf A2} and {\bf A3} can be, similarly,
assigned to one of the three halos ({\bf P1}, {\bf P2}, and {\bf P3}).


\subsection{Tidal Radius Criterion}
\label{trsec}
The \hfind method uses multiple steps to assign particles to halos.
Particle sets are sorted in a decreasing order of density levels.
Starting from particles of the top inner-most PS,
we
apply the total energy check to each particle and 
the tidal radius criterion to each halo enclosed by the PS.
But there needs an iterative scheme in this hierarchical membership determination.
Since a halo mass is usually measured by its member particles,
a halo potential may be underestimated
unless all PS's down to the bottom one are taken into account.
To supplement the halo mass, 
we use the PS's having the density levels both higher and lower than that of  
the current particle set $\rm PS_c$.
Let the $\PS_{\downarrow}$ be the particle sets geometrically enclosing the halo
and having density levels lower than that of the $\rm PS_c$.
The potential $\Phi(k)$ of the $k$'th particle 
with respect to its host halo is calculated by
\begin{equation}
\Phi(k\in {\rm PS_c}) = -\sum_{\ell \in MEM} {Gm_{\ell} \over r_{k\ell}} - 
\sum_{\ell\in \PS_{\downarrow}} {Gm_{\ell} \over r_{k\ell}},
\label{pot}
\end{equation}
where $r_{k\ell}=| \vec r_{\ell} - \vec r_k |$,
$m_{\ell}$ is the $\ell$'th particle mass, 
$MEM$ are the member particles already assigned to the halo
in the $\PS_{\uparrow}$'s which have density thresholds higher than 
that of the $\rm PS_c$.
The second term of the right hand side of equation (\ref{pot})
significantly contributes to the particle potential $\Phi$
if the number of particles in $\PS_{\downarrow}$'s is large.
We do not include particles of the $\rm \PS_c$ for the halo mass
in equation (\ref{pot})
since an accidentally close particle to the $k$'th particle
can make the potential overestimated.

In some cases, particles happen to be bound to many halos simultaneously.
To resolve this situation,
we propose a new halo boundary indicator, the tidal radius, which is based
on the local gravitational field.
A tidal radius of a halo is calculated 
from a tidal mass $M_t$ which is the number of only the halo member particles
\begin{equation}
M_t = \sum_{k\in MEM} m_k.
\label{mt}
\end{equation}
Let $R_t(i,j)$ be the tidal radius $R_t(i)$ of the $i$'th halo 
with respect to the $j$'th halo more massive than the $i$'th.
A simple analytic form for the tidal radius is the Jacobi radius $R_J$
\begin{equation}
R_J(i,j) \equiv D \left( {M_t(i)\over 3M_t(j,r<D)}\right)^{1/3},
\label{kingtidal}
\end{equation}
where $D$ is the distance between the centers of mass of the halos,
and $M_t(j,r<D)$ is the tidal mass of 
the $j$'th halo mass contained in a sphere with a radius $D$.
The effect of underestimation of the  halo mass (Eq. \ref{mt})
on the tidal radius (Eq. \ref{kingtidal}) is thought to be insignificant
due to three reasons.
First, $R_J$ is proportional to its cubic root.
Secondly, we measure $M_t$ at each PS analysis and
iteratively update the tidal radius of each halo.
Then the tidal radius of a halo converges to a true value when the bottom PS
is considered.
Thirdly, the $j$'th halo member particles beyond $D$
do not contribute both to $M_t(j,r<D)$ and $M_t(i)$.
The particles of a ${\PS}_\downarrow$ that surrounds both the $i$'th 
and $j$'th halos
are usually too far from the two halos to contribute to the tidal masses.
And it is less likely for the particles of the ${\PS}_\downarrow$ to be
assigned to the $i$'th halo and, $M_t(i)$ is not expected to increase
significantly.
Therefore, even if we initially 
assign an inaccurate value to the tidal radius,
the tidal radius of each halo converges to a true value
in this hierarchical particle assignment process.

\begin{figure}[tbp]
\begin{minipage}[t]{7cm}
\leavevmode
\epsfxsize=8cm
\epsfysize=8cm
\epsffile{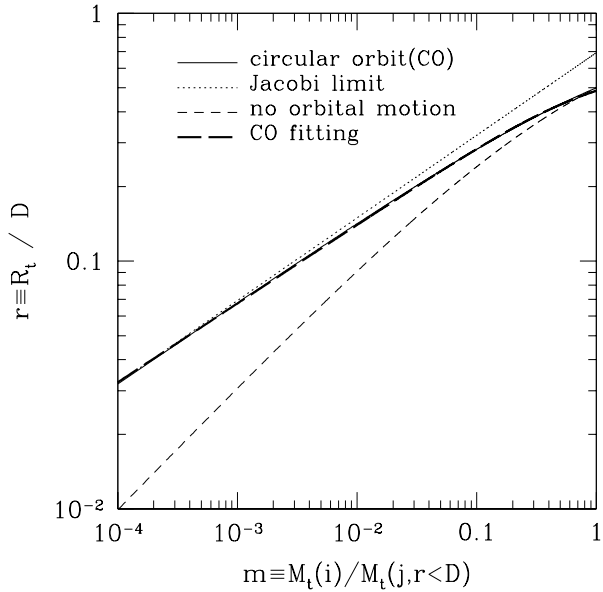}
\end{minipage}
\caption
{$\mathfrak{r}$-$\mathfrak{m}$ relation.
We show the ratio $\mathfrak{r}$ of the tidal radius to the distance between halos
as a function of the mass ratio between $M_t(i)$ and 
$M_t(j,r<D)$ for the circular orbit (solid curve), 
for the Jacobi limit (dotted curve, Eq. \ref{kingtidal}),
and for the system with no orbital motion (short-dashed curve).
The long-dashed curve shows the fitting result
of the tidal radius for a circular orbit system adopted in the \hfind method.
}
\label{tidalfig}
\end{figure}

However, the Jacobi radius $R_J$ (Eq. \ref{kingtidal}) 
is obtained assuming that $M_t(i)/M_t(j,r<D) \ll 1$.
To take into account more general situations,
we calculate the tidal radius $R_t$ of a smaller halo
as a function of the halo mass ratio 
$\mathfrak{m} \equiv {M_t(i)/M_t(j,r<D)}$ using
\begin{equation}
\left(1-\mathfrak{r}\right)^{-2} - 
\mathfrak{m}\mathfrak{r}^{-2} - 1
+ \left( 1+\mathfrak{m}\right) \mathfrak{r} = 0,
\label{trsol}
\end{equation}
where $\mathfrak{r}\equiv R_t/D$ (Eq. 7-82 of Binney \& Tremaine 1994). 
We assume that two halos orbit circularly
around their center of mass.
In Figure \ref{tidalfig},
we show the ratio of the tidal radius $R_t$ to the distance $D$
as a function of the halo mass ratio $\mathfrak{m}$
under various conditions such as the circular orbital motion
(Eq. \ref{trsol}, solid curve),
the Jacobi limit $\mathfrak{m}\rightarrow0$ 
(dotted curve, King 1962; Binney \& Tremaine 1994),
and no orbital motion (short-dashed curve).
As can be seen in Figure \ref{tidalfig}, our adopted tidal radius model
(solid curve) can be approximated by the Jacobi radius $R_J$ in the limit of
$\mathfrak{m} \rightarrow 0$.
However, as $\mathfrak{m}\rightarrow1$ in a binary system with equal halo masses, 
the Jacobi radius $R_J$ approaches an incorrect value of 0.69$D$
while our formula gives $R_t=D/2$.
We fit our model to a fitting function (Keenan 1981)
\bea
R_t = D \left[ {M_t(i)\over \alpha\left(M_t(i)+M_t(j,r<D)\right)}\right]^{\beta},
\label{tidalr}
\eea
where $\alpha=4.813$ and $\beta=0.318$ (long-dashed curve).

To a small halo many tidal radii can be assigned with respect to
other more massive halos enclosed by a given PS.
We adopt the minimum value of $R_t(i)$'s as
\begin{equation}
R_t(i) ={\rm MIN}[R_t(i,j=1),\cdots,R_t(i,j=N_{eh})], 
\label{mintidal}
\end{equation}
where $N_{eh}$ is the number of enclosed halos 
with mass satisfying $M_t(j,r<D_{ij}) > M_t(i)$,
where $D_{ij}$ is the distance between the $i$'th and $j$'th halos.
If a particles is not bound to any halo or
beyond the tidal radius of each halo,
we stack it in the free-particle list.
Particles in the free-particle list
are temporarily added to the particle list of 
the next PS.

After completing the above steps for all $\PS$'s
we use the Friend-of-Friend (FoF) method
with $l_{\rm FoF} = 0.2 \times l_{\rm mean}$
to find particles belonging to virialized halos.
Particles which are not members of any halo 
are stacked in the free-particle list again.

Before we get final halo data,
we check whether or not halos are self-bound and virialized.
This process is similar to those of other halo finding methods
except for the tidal radius criterion.
We find the minimum tidal radius of each halo using equation (\ref{mintidal}).
At this time, 
$N_{eh}$ is the number of all halos more massive than a given halo
in the local-cell group region.
We look into the member-particle and free-particle lists.
Unbound particles or those beyond the tidal radius of a halo are discarded and
stacked to the free-particle list.
We iterate this process four times to obtain self-bound and stable halos.
Finally, the FoF method is applied to satisfy the virialization condition
of each halo.

\begin{center}
\begin{figure}[tbp]
\begin{center}
\begin{minipage}[t]{7cm}
\leavevmode
\epsfxsize=7cm
\epsfysize=15cm
\epsffile{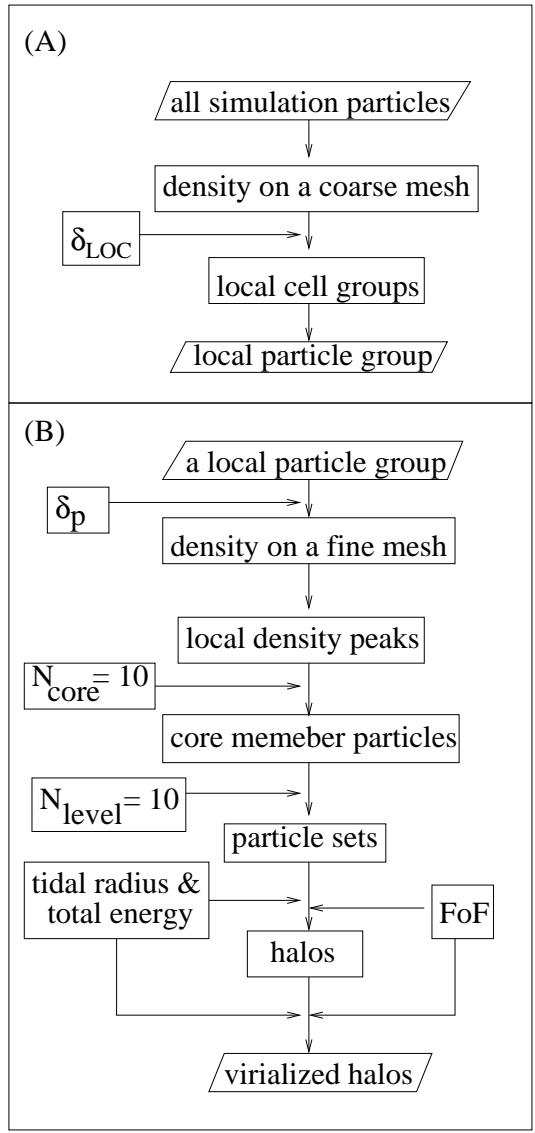}
\end{minipage}
\end{center}
\caption
{
Flowchart of the \hfind method.
We show the way to extract local particle groups
based on a coarse mesh described in panel (A), and
to identify virialized halos
from a local particle group in panel (B).
There are four critical parameters used in the PSB method,
such as $\delta_{\rm LOC}$, $\delta_p$, $N_{core}$, and $N_{level}$.
}
\label{flowchart}
\end{figure}
\end{center}
The PSB method is summarized in the flowchart in Figure \ref{flowchart}.
In panel (A), we show the pre-halo finding process described in \S\ref{ilgg}.
To find local particle groups, we calculate the density on a low-resolution
mesh using all simulation particles. Using densities on the mesh
and a predefined parameter $\delta_{\rm LOC}$,
we find overdense cells and link them making local-cell groups.
Particles on the linked cells are grouped and extracted forming local particle
groups.
Using a local particle group, the PSB method identifies virialized halos.
Panel (B) shows a flowchart of the halo finding process on a fine mesh.
After finding particle sets, we iteratively update the tidal radius of a halo
and calculate the total energy of member particles.
We, then, apply the FoF method to find member particles which 
satisfy the virialization condition.



\subsection{Parallelization}
\label{hfparallel}
We parallelize our halo finding method using the Message-Passing-Interface
(MPI) library.
As described previously, the \hfind method is based on 
two meshes with coarse and fine cells.
It therefore adopts two different modes of domain decomposition.
Since a coarse mesh is built over the entire simulation box,
it is desirable for each computational rank to have the same domain size.
We use the domain decomposition into the z-directional
domain slabs as described in the GOTPM code (Dubinski et al. 2004).
We search for overdense cells and connect them together.
Overdense cells at the upper and lower boundaries of local domains
can be linked to neighboring domains.
Because we use one-dimensional cyclic ordering of local domains,
we pad images of mesh slices imported from neighboring domains 
at the upper and lower boundaries of the local domain.
Those imported mesh slices contain local informations of their domains.
We also take account of the periodic boundary conditions of the domain slab 
in the horizontal directions.
With these linked groups of cells, we pick up particles 
in the overdense cell regions and save those particles to disk.

On the fine mesh the domain decomposition is made based on the local particle group.
We set a master rank to do the disk I/O of the local particle data
and to distribute them into one of the slave rank that is idle at that time.
Each slave rank receives particle data and performs the process
described in the \S\ref{denf} and \S\ref{trsec}.
After finishing the halo finding procedure in a local particle group,
the slave rank sends halo member particles back to the master rank
to save the halo finding results.

%

\section{Parameter Determination}
\label{vdeltal}
The lower $\delta_{\rm LOC}$ is, the smaller are the smallest halos identified.
Two important factors constraining $\delta_{\rm LOC}$
are the computer memory available for each rank and 
the maximum allowable number of density peaks in a local particle group.
Since a fine mesh devours much memory budget,
the mesh with about $750^3$ cells is the maximum allowable size
on 32bit machines.
Second, the tidal radii should be measured for
a larger number of density peaks
when there are many density peaks in a local particle group.
Counting particles to measure $M_t(j,r<D)$ in eq. (\ref{tidalr})
with respect to every smaller halo consumes a lot of CPU time.

We have empirically found that it is adequate to set
$\delta_{\rm LOC} = 1$--$3\sigma_2(0)$, namely,
$\delta_{\rm LOC}\sim \sigma_2(0)$ for the case of the SCDM
and $\delta_{\rm LOC}\sim 3\sigma_2(0)$ for 
the case of the concordance $\rm \Lambda$CDM model where
$\sigma_2(0)$ is the standard deviation of density fluctuation 
measured in the coarse mesh at $z=0$.
We use a constant value of $\delta_{\rm LOC}$ at all redshifts.
A lower value of $\delta_{\rm LOC}$ enable one to identify smaller halos
in the lower density regions, but the volume of each local-cell group
increases significantly.
In the extreme case of $\delta_{\rm LOC}=-1$,
there is one big local-cell group containing all simulation particles.
Therefore, the advantage of dividing all simulation particles into
multiple local particle groups is no more sustained.
On the other hand, higher value of $\delta_{\rm LOC}$ is not desirable 
since the tidal radii can be underestimated 
due to the insufficient description of the local tidal field.
Additionally, small halos residing in lower dense regions
can be missed in the process of the halo findings.
We choose larger $\delta_{\rm LOC}$ for the $\rm \Lambda$CDM model
because the \LCDM model 
has higher degree of connectivity of overdense regions
(filamentary structures) than the SCDM model.
Above values of $\delta_{\rm LOC}$ are determined for 2GB memory budget.
If more memory budget is available,
we are able to reduce $\delta_{\rm LOC}$ and identify smaller halos
in lower dense regions.

To avoid detecting spurious density peaks, 
we require density peaks to have at least two particles
in the fine mesh cell.
The density corresponding to this constraint is
\begin{equation}
\delta \sim 2\times (l_{\rm mean}/l_{f})^3,
\end{equation}
where $l_{f}$ is the cell size of the fine mesh.
If $l_{\rm mean}/l_{f}=5$, then $\delta\sim 250$.
A halo having the density peak cell with two particle
satisfies the virial condition $\delta_{vir}\ge177$ automatically.
Two density parameters $\delta_{\rm LOC}$ and $\delta_p$
should satisfy the virial condition,
\begin{equation}
\delta_{\rm LOC} \ll \delta_{vir} < \delta_p.
\end{equation}

In the \hfind method, two density meshes with different cell sizes
are adopted and different density thresholds, $\delta_{\rm LOC}$ and
$\delta_p$, are applied to them.
Occasionally, there may be peaks with $\delta > \delta_p$ at the scale of
the fine cell but with $\delta<\delta_{\rm LOC}$ on the coarse cell.
We estimate the fraction of such missing peaks in the following
to show the problem is minor 
for our choice of $\delta_{\rm LOC}$ and $\delta_p$.

For a Gaussian density field,
the probability of finding the regions with $\delta \ge \delta_t$ 
is
\begin{equation}
P[\delta\ge\delta_t] = {1\over \sqrt{2\pi}\sigma}
\int_{\delta_t}^\infty d\delta^\prime
\exp{\left({ {\delta^\prime}^2 \over 2 \sigma^2}\right)},
\end{equation}
where $\sigma^2$ is the variance of the density fluctuation.
Let us use the subscript 1 to denote the statistical quantities measured
in the fine mesh and the subscript 2 in the coarse mesh.
In our halo finding method the smoothing lengths $h$'s 
are set to the mesh cell sizes, namely
$h_1 = l_1 = 2 \epsilon$ and $h_2=l_2=l_{mean}$.
It is usually $\epsilon = 0.1 \times l_{\rm mean}$ 
in high resolution cosmological simulations.
The conditional probability that a region has $\delta \ge \delta_p$
in the fine mesh when it has $\delta\le\delta_{\rm LOC}$ 
in the coarse mesh is,
\bea
\nonumber
&P&\left[\delta_{h_1}\ge\delta_p| \delta_{h_2}\le\delta_{\rm LOC}\right]  \\
&=& P\left[\delta_{h_1}\ge\delta_p, \delta_{h_2}\le\delta_{\rm LOC}\right] /
P\left[\delta_{h_2} \le \delta_{\rm LOC}\right].
\eea
The joint probability $P\left[\delta_{h_1}=\delta_1, \delta_{h_2}=\delta_2\right]$ is
\bea
\nonumber
&P&\left[\delta_{h_1}=\delta_1, \delta_{h_2}=\delta_2\right] \\
\nonumber
&=& 
{1\over {2\pi \sqrt{|M|}}} \exp\left({-{1\over2}\Delta^TM^{-1}\Delta}\right) \\
\nonumber
&=& {1\over \sqrt{2\pi}\sigma_s} \exp\left[-{1\over2}
{\left({\delta_1-{\delta_2\sigma_c^2\over\sigma_2^2}}\right)^2/\sigma_s^2}
\right]\\
&\times& {1\over \sqrt{2\pi}\sigma_2} \exp\left(-{1\over 2} { \delta_2^2\over \sigma_2^2}\right)
,
\eea
where
\bea
\Delta &\equiv& { \delta_1 \choose \delta_2}, ~~~
M \equiv \left(\begin{array}{c c}
\sigma_1^2 & \sigma_c^2 \\
\sigma_c^2 & \sigma_2^2
\end{array}\right),\\
\sigma_i^2 &=&{1\over (2\pi)^3} \int d^3k P(k) |{W_{h_i}}|^2,\\
\sigma_c^2 &=&{1\over (2\pi)^3}
\int d^3k P(k) \tilde{W}_{h_2}^\ast(k)
\tilde{W}_{h_1}(k),  \\
\sigma_{s}^2 &=& \sigma_1^2 - {\sigma_c^4\over\sigma_2^2},
\eea
and $P(k)$ is the power spectrum of the density field.
The smoothing kernel $\tilde{W}_k$ in $k$-space is given by
\bea
\tilde {W}_h({\bf k}) = {1\over (2\pi)^3} 
\int_0^\infty dr^3 W_4({\bf r};h) \exp{\left(i {\bf k}\cdot {\bf r}\right)},
\eea
and we use a spherically symmetric smoothing window $W_4$
(Monaghan \& Lattanzio 1985).
Then, we get
\bea
\nonumber
&P&\left[\delta_{h_1}\ge\delta_p| \delta_{h_2}\le\delta_{\rm LOC}\right] \\
\nonumber
&=&
{1\over{2\pi}\sigma_2\sigma_s N}
\int_{-1}^{\delta_{\rm LOC}} d\delta_2 
\int_{\delta_p}^\infty d\delta_1 
\exp\left(-{\delta_2^2\over2\sigma_2^2}\right) \\
&&
\cdot \exp\left[- {\left({\delta_1 -{\sigma_c^2\over\sigma_2^2}\delta_2}\right)^2
/2\sigma_{s}^2}\right],
\label{probability}
\eea
where 
\bea 
\nonumber
N&\equiv& P\left[\delta_{h_2}\le\delta_{\rm LOC}\right]\\
&=& {1\over{\sqrt{2\pi}\sigma_2}}
\int_{-1}^{\delta_{\rm LOC}} d\delta_2 
\exp\left({- {\delta_2^2\over{2\sigma_2^2}}}\right).
\eea

We have applied our halo finding algorithm to two cosmological N-body 
simulation data.
The simulations are based on a $\rm \Lambda$CDM 
and the SCDM models with the same simulation box size
$L_{box}=512h^{-1}{\rm Mpc}$.
In the \LCDM and the SCDM models
we follow the gravitational evolution of 
$1024^3$ particles in $1024^3$ mesh cells from $z=23$ to $z=0$
with 460 time steps by using the GOTPM code (Dubinski et al. 2004).
We set $\delta_{\rm LOC}=10$ ($\simeq 3\sigma_2(0)$) for the \LCDM model and 
$\delta_{\rm LOC}=5$ ($\simeq \sigma_2(0)$) for the SCDM model,
and fix the peak density threshold $\delta_p= 312.5$.
We calculate the number of unidentified density cells ($\delta_1 >\delta_p$) 
which reside in the underdense background ($\delta_2 < \delta_{\rm LOC}$)
using equation (\ref{probability}).
At $z=0$ the number of such density peaks on the fine mesh
in the total simulation box is only a few,
which is very small compared to the number of halos ($\sim 2\times 10^6$
in the \LCDM simulation and $\sim 4\times 10^6$ in the SCDM simulation)
identified by the halo finding method.

In some situations, 
small halos identified in overdense backgrounds
may have the true peak densities lower than $\delta_p$
when the underlying backgrounds are removed. To obtain unbiased halo samples
we calculate the halo density by using their
member particles at the end of halo identification,
and discard the halos having the peak density $\delta <\delta_p$ from the halo catalog.
Finally, for our \LCDM and SCDM simulation data
we obtain halos with masses larger than $6\times10^{11}h^{-1}{\rm M_\Sun}$
($\simeq 58$ particles) and
$10^{12}h^{-1}{\rm M_\Sun}$ 
($\simeq 30$ particles) , respectively.

\section{Sensitivity to Parameters}
\label{parasec}

Using halo mass functions, we examine the dependence of halo finding results
on the four group finding parameters,
$\delta_{\rm LOC},\delta_p^\prime,N_{level}$, and $N_{core}$,
where $\delta_p^\prime\equiv\delta_p/125$.
As a testbed, a \LCDM model is simulated with $128^3$ particles in a box
of size $128h^{-1}\MPC$.
The bias factor is set to $b=1.11$ and the initial epoch is $z_i=4$.
At $z=0$, the RMS of linear density fluctuation on the coarse mesh is 2.69.
We use the canonical parameter set 
$(\delta_{\rm LOC},\delta_p^\prime,N_{level},N_{core}) =(11,2.5,10,10)$.
Then $\delta_p^\prime\simeq4.1\times \sigma_2(0)$.

\begin{figure}[tbp]
\begin{minipage}[t]{ 8cm}
\leavevmode
\epsfxsize=8cm
\epsfysize=8cm
\epsffile{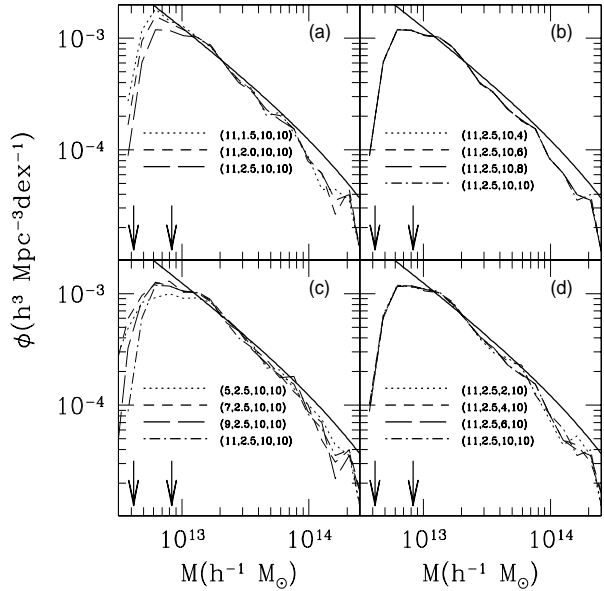}
\end{minipage}
\caption
{
Halo mass functions with various different parameter sets.
The canonical parameter set is obtained
$(\delta_{\rm LOC},\delta_p^\prime,N_{level},
N_{core}) = (11,2.5,10,10)$, where $\delta_p^\prime\equiv\delta_p/125$.
In panel (a), we show the halo mass functions varying $\delta_p^\prime$
from 1.5 to 2.5.
Panel (b) is same as panel (a), but $N_{core}$ is varied.
We vary $\delta_{\rm LOC}$ in panel (c) and
$N_{level}$ in panel (d).
The solid line shows the Sheth \& Tormen mass function,
and the two vertical arrows mark the mass of 50 and 100 particles.
}
\label{para}
\end{figure}
In panel (a) of Figure \ref{para}, we show 
halo mass functions with varying $\delta_p^\prime$ from 1.5 to 2.5.
When lower values of $\delta_p$ are adopted,
much more smaller halos are found.
In panel (b)
the halo mass function shows negligible dependence on $N_{core}$.
As the peak parameters,  $\delta_p$ and $N_{core}$ are correlated 
with each other.
The variation of $N_{core}$ apparently 
takes no effect on the halo mass functions since $\delta_p$ 
is a more stringent parameter
when $\delta_p$ is high enough to satisfy $\delta_p \ge \delta_{vir}$.
The dependence of the halo mass function on the threshold 
density $\delta_{\rm LOC}$ is also weak as shown in panel (c).
The parameter $N_{level}$ hardly affects
the halo mass functions as demonstrated in panel (d).
We use a high value of $N_{level}=10$ to better resolve clouded regions
with many peaks.
For example, there are about two thousands peaks 
in the largest local particle group 
in our $1024^3$ particle simulations at $z=0$.

We conclude that the most important parameter is $\delta_p$ 
which determines the lower limit of mass of the halos found. 
The role of $\delta_{\rm LOC}$ in the \hfind method
is less significant when $\delta_p$ is high.
But if one wants to identify less massive halos or to reduce $\delta_p$, 
attention must be paid to the value of $\delta_{\rm LOC}$.
He must measure 
the probability of missing fine cells which have $\delta \ge \delta_p$
in the underdense background ($\delta \le \delta_{\rm LOC}$).
The other parameters are proved to be insensitive to 
the halo finding results in the \hfind method.

\section{Comparison With Other Halo Finding Methods}
\label{hfcomp}
\begin{figure}[tbp]
\begin{minipage}[t]{8.3cm}
\leavevmode
\epsfxsize=8.3cm
\epsfysize=13cm
\epsffile{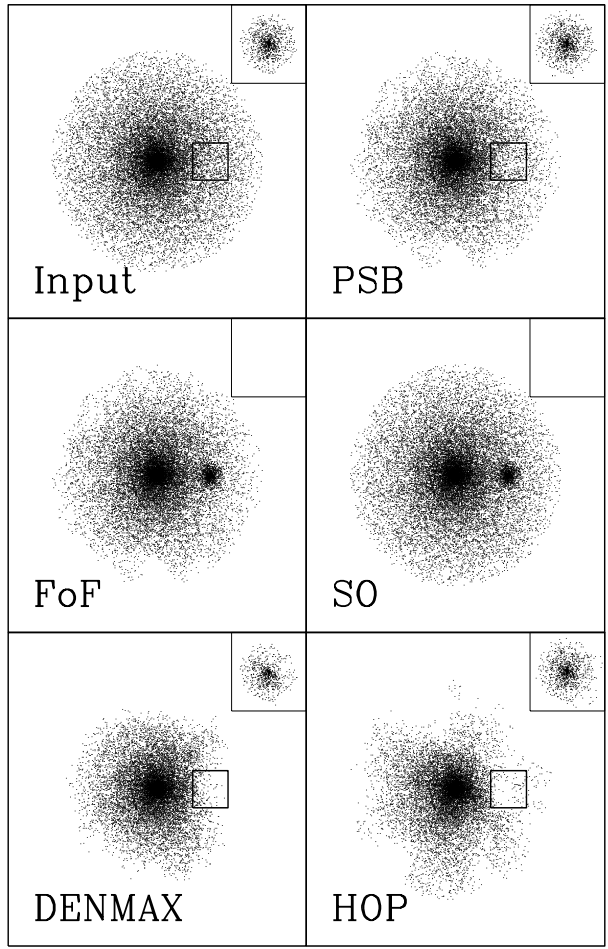}
\end{minipage}
\caption
{
Comparison among the results from various halo finding methods.
We show the model distribution of particles of two halos
in the upper-left panel.
The large halo has 20000 particles following the isothermal density profile.
The small halo whose region is marked by a box has 1000 particles 
with the isothermal but tidally truncated density slope.
In the inset, we show the enlarged view of the small halo for clarity.
The small halo is offset to right with respect 
to the large halo.
We show the identified particle distribution obtained by
halo finding methods,
such as the FoF (middle-left panel), SO (middle-right panel),
DENMAX (lower-left panel), HOP (lower-right panel), and
\hfind method (upper-right panel).
The insets at the upper-right corners 
show the small halos identified by these methods.
}
\label{ohf}
\end{figure}

We have compared the results of the PSB method with
those of other competing halo finding methods.
We construct a binary system of halos having 20000 and 1000 member particles,
respectively.
The large halo has the virial radius 
$R = (3M_h/4\pi\bar{\rho}_{178})^{1/3}$,
and the small halo has its tidal radius
under the tidal field of the large halo.
The distribution of the member particles is set to follow 
the isothermal density profile.
The small halo is slightly offset in x-direction 
(upper-left panel in Fig. \ref{ohf}).
The directions of particle velocities are chosen randomly
and the magnitudes are given in accordance with 
the virialization condition.
Here, the small halo is bound to the potential well of the large halo,
with no bulk motion.
The upper-left panel of Figure \ref{ohf} shows the particle distribution of 
each halo with the enlarged particle distribution of the small halo in the inset.
We also mark the small halo region with a box in the panel but 
do not overplot the member particles of the small halo for clarity.
The middle-left panel shows the result of the FoF method
which does not identify the small halo.
The halo boundary found by the FoF method is not smooth
due to the Poisson fluctuation of
particles at the outskirts of the large halo.
The SO method cannot identify the small halo like the FoF method.
The large halo is found with a spherical boundary, which is a
characteristic of the SO method.
These two methods cannot detect the small halo since the halo identification
is mainly based on the mean overdensity criterion.

The lower-left panel shows the result of the DENMAX method,
which resolves the small halo.
However, due to the sliding of particles to the nearby densest cell
the large halo seems to have no member particles at the right hand side
of the small halo region.
All those particles are assigned to the small halo, but 
they are not bound because of the shallow potential well of the small halo.
Consequently, those particles are regarded unbound, and 
erased from the member list of the small halo. Having the
particle sliding algorithm similar to that of the DENMAX,
the HOP method can detect particles beyond the small halo region.
Those particles can slide according to local particle links,
and possibly make a detouring path around the small halo region 
to the density maxima of the large halo.
Since we do not use the REGROUP algorithm
to regroup particles stacked in the local and small density maxima,
there are many ``particle holes'' in the large halo 
identified by the HOP result.

The upper-right panel shows the result of our \hfind method.
The boundary of the large halo is similar to that in the FoF method
because the FoF method is used in our algorithm 
to find member particles in the virialized region.
The \hfind method identifies member particles of the small halo
nearly the same as the input.
For examples, the numbers of identified member particles of the small halo are
0, 0, 758, 1101, and 1091 for
the FoF, SO, DENMAX, HOP, and \hfind methods , respectively. 
Therefore, the \hfind method seems to recover 
the input particle distributions more accurately than other methods
except for the HOP.
But the density allocation method of the HOP requires more CPU time 
to search for $N_{neighbor}$ particles 
than the \hfind method which uses a fixed smoothing length.
We conclude that the \hfind method is an efficient and accurate
halo finding method that can be adapted to
the parallel computing environments.

\section{Result of Application}
\label{hfresult}

\subsection{Halo Finding for 1-Giga particle simulations}
\begin{figure}[tbp]
\begin{minipage}[t]{ 8cm}
\leavevmode
\epsfxsize=8cm
\epsfysize=8cm
\epsffile{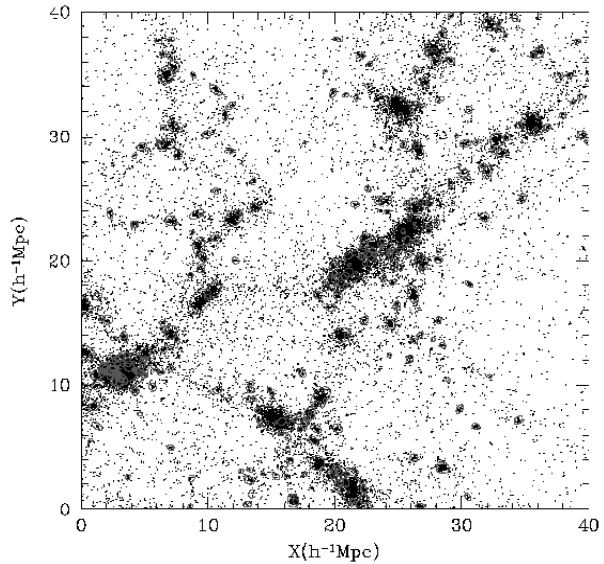}
\end{minipage}
\caption
{
Projected particle distribution in the \LCDM simulation.
The thickness of the slice is $10h^{-1}\MPC$.
Only every tenth particles in the volume are shown for clarity.
Projected ellipsoidal shapes (marked by ellipses)
of total 355 halos identified in this volume are shown.
}
\label{projectionhf}
\end{figure}
We apply our halo finding method to the 1-Giga particle data simulated 
on the IBM SP3 
at Korea Institute of Science and Technology Information (KISTI) 
and on a Linux cluster.
The times needed to complete the halo findings at $z=0$
in the SCDM and the \LCDM models are about four hours 
when 32 POWER-4 CPUs are used.
In Figure \ref{projectionhf}, we show the result of halo finding 
in the \LCDM model at $z=0$.
Only every tenth particles in the volume are shown for clarity.
The projected ellipsoidal shapes of halos
are plotted on top of the particle distribution.

\subsection{Mock Galaxy Redshift Survey}
\begin{figure}[tbp]
\begin{minipage}[t]{ 8cm}
\leavevmode
\epsfxsize=8cm
\epsfysize=8cm
\epsffile{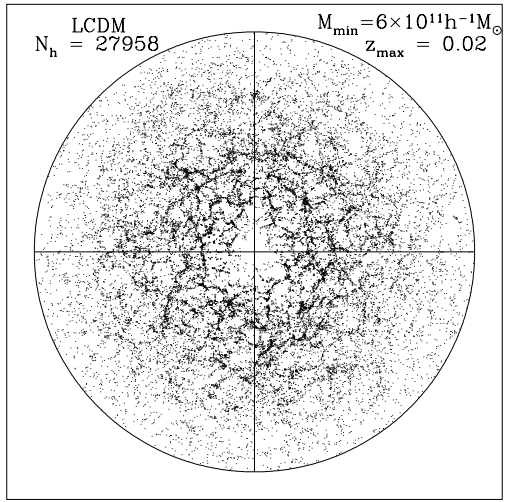}
\end{minipage}
\caption{
A mock redshift survey in the \LCDM model mimicking the LCRS.
The simulated survey slice is $360^\circ$ by $4.5^\circ$ wide.
Total number of mock galaxies  is 27958,
with mass $M\ge 6 \times 10^{11}h^{-1}{\rm M}_\Sun$.
The highest redshift $z_{max}$ is 0.02.
}
\label{mocklcdm}
\end{figure}

Figure \ref{mocklcdm}, 
shows the distribution of halos in a mock galaxy redshift survey 
similar to the Las Campanas Redshift Survey (hereafter LCRS, Lin et al. 1996).
The selection function adopted here is the same as that of the LCRS.
We do not take into account the conversion of halo mass functions to 
galaxy luminosity functions (Yang et al. 2003).  
The effects of peculiar velocities of halos are included.
The halo number density in our \LCDM simulation is 
$\rho_h=0.0082h^3/{\rm Mpc}^3$
which is about 3.5 times less than that 
($\rho_g = 0.029\pm0.002h^3/{\rm Mpc}^3$)
of the LCRS galaxies in absolute magnitude range 
$-23.0 \le {\cal M}-5\log h \le -17.5$.
Using the Schechter function 
\bea
\nonumber
\phi({\cal M}) &=& (0.4\ln 10) \phi^\star\left[10^{0.4\left({\cal M}^\star
-{\cal M}\right)} 
\right]^{(1+\alpha)} \\
&\times& \exp\left[ -10^{0.4\left( {\cal M}^\star-
{\cal M}\right)}\right],
\eea
where ${\cal M}^\star=-20.29$,
$\alpha=-0.7$, and $\phi^\star=0.019$ (Lin et al. 1996),
we find that the mass $6\times 10^{11}h^{-1}{\rm M}_\Sun$ corresponds to
the magnitude $-19.8+5\log h$ in the LCRS galaxies
from a direct comparison between the
cumulative number density of the LCRS galaxies and 
that of the \LCDM halos.

\subsection{Isolated and Distinguished Halos; Halo Mass Function}
In this subsection, we compare the halo mass functions found by the \hfind
and the FoF-like methods with the well-known analytic and fitting functions.
We call the halo found by the \hfind the distinguished halo.
A distinguish halo is a virialized structure irrespective of 
its environment.
For example, each halo in Figure \ref{projectionhf}
is regarded as a distinguished halo.
The FoF-like halo catalog is constructed by merging halos
when halo regions are overlapped.
To model each halo as an ellipsoid
we use the following shape tensor 
\begin{equation}
\mathrm{M}_{ij} = \sum_{k=1}^N m_k x_i(k) x_j(k), ~~~~ i,j=1,2,3,
\label{shapetensor}
\end{equation}
where $m_k$ is the mass of the $k$'th member particle 
and $x_i$ is its position with respect to the halo center of mass.
The shape tensor $\mathbf{M}$ is diagonalized for each halo.
The ratios of principle axes are the square roots of ratios of the eigenvalues,
and the orientations of the axes are the corresponding eigenvectors.
However, 
the volume of the halo cannot be obtained from equation (\ref{shapetensor}).
Conserving the ellipsoidal shape of the halo, we scale up and down the shape
to include almost all member particles.
During the scaling, we set the virial volume to include
95\% of member particles since the halo boundary is not smooth.
We check whether or not 
the virial region of a halo overlaps with those of other halos.
If overlapping with each other, 
those halos are merged into a single halo.
This merging process is similar to the linking of the FoF method.
A halo found by this FoF-like method is a spatially isolated system 
whose virial region is not shared with other halos.
We call this halo an isolated halo.

\begin{figure*}[tbp]
\begin{center}
\epsfxsize=15cm
\epsfysize=12cm
\epsffile{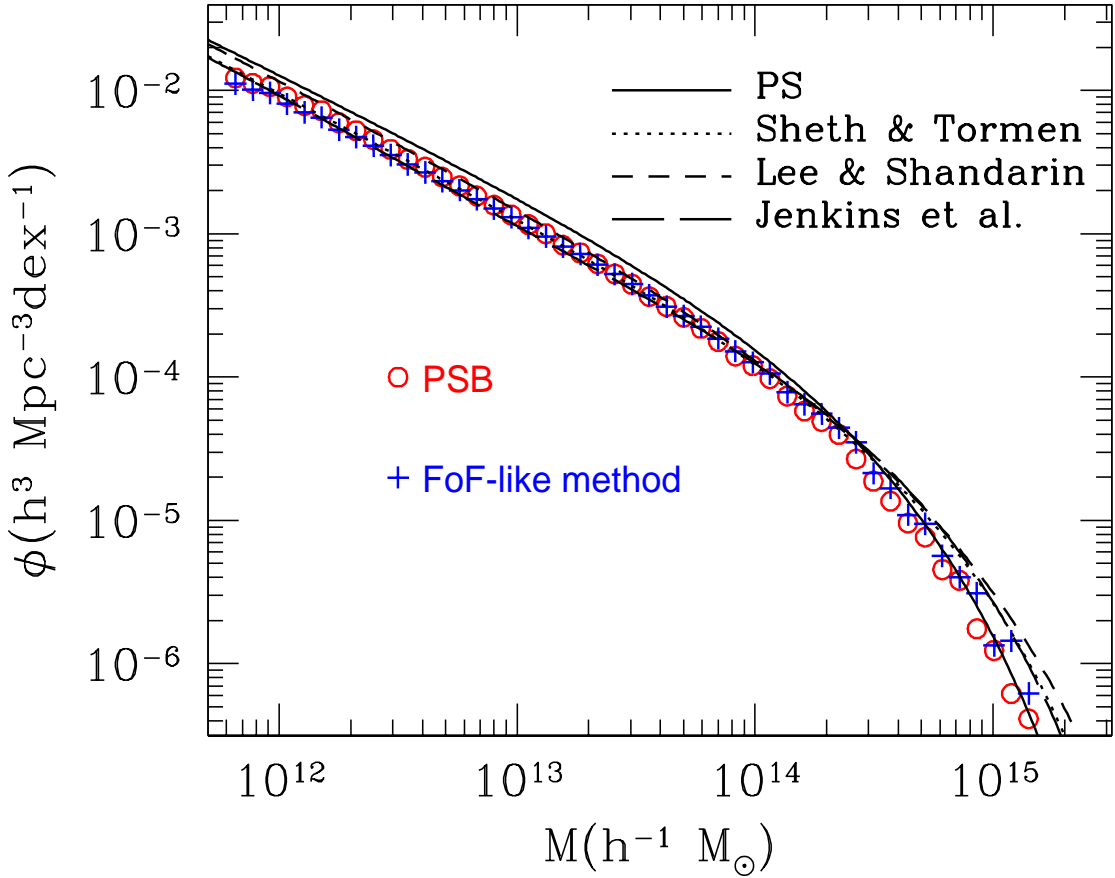}
\end{center}
\caption{
Halo mass functions obtained from the \LCDM model simulation at $z=0$.
The halo mass functions obtained by the \hfind (open circles)
and the FoF-like method (crosses) are shown.
We also plot the analytic and fitting functions
proposed by Press \& Schechter (solid curve),
and Sheth \& Tormen (dotted curve), Lee \& Shandarin (short-dashed curve),
and Jenkins et al. (long-dashed curve).
}
\label{foflocpg}
\end{figure*}
In Figure \ref{foflocpg}, we show the halo mass functions of the \LCDM model
simulation at $z=0$ measured from the distinguished halo (open circles)
and the isolated halo catalogs (crosses).
Also shown are other analytic and fitting functions, such as 
the Press \& Schechter (1974, solid curve), 
Sheth \& Tormen (1999, dotted curve),
Lee \& Shandarin (1998, short-dashed curve), 
and Jenkins et al. (2001, long-dashed curve) mass functions.
In the mass range of $M_h\le2\times10^{14}h^{-1}{\rm M_\Sun}$, 
the FoF-like method underestimates the number of halos, 
and overestimates the number of halos with mass
$M_h>2\times10^{14}h^{-1}{\rm M_\Sun}$ when compared with the \hfind catalog.
Small distinguished halos embedded in dense regions are merged into large halos
and, as a result, the mass function of isolated halos approximately follows
the predicted mass function of Jenkins et al. (2001) who used the FoF method 
to find isolated halos in the Hubble Volume Simulations (Colberg et al. 2000).
We will study the this isolated and distinguished halos
in the forthcoming paper in more details.

\section{Conclusions}
\label{hfcon}
We provide a new halo finding algorithm that efficiently and
accurately identifies halos located in dense environments.
This method employs a tidal radius constraint as well as the total energy criterion
in determining the halo membership .
Based on the density field on a fine mesh with cell size of $l_1 = 2\epsilon$,
we detect halos with core size down to the force resolution $\epsilon$.
We have compared the \hfind method
with other popular halo finding methods 
in resolving small halos embedded in a larger halo.
The FoF and SO methods fail to identify the small halos.
The tidal radius of a halo is 
a physically meaningful boundary descriptor, particularly in crowded regions.
Density-related descriptors are not enough to identify halos
when they are overlapping each other.
When two halos are close to each other, 
the halos found by the HOP and DENMAX methods
may have
wrong physical characteristics such as the total mass, shape, spin,
peculiar velocity and etc.
The adaptive smoothing of the HOP method
is not consistent with N-body simulations since the gravitational smoothing
length $\epsilon$ is kept constant throughout whole simulation.
It is more reasonable to have 
the smoothing length $h$ of the interpolating kernel $W_4$
equal to $\epsilon$.
The \hfind algorithm is faster than the HOP
because of the static smoothing length in obtaining the density field.

The \hfind method has several parameters.
The most important ones are $\delta_{\rm LOC}$ and $\delta_p$.
We use the constraint of $\delta_{\rm LOC} \ll \delta_{vir}< \delta_p$ 
to find virialized halos with peak density greater than $\delta_p$ and 
to accurately calculate the tidal effect in the local overdense regions 
($\delta>\delta_{\rm LOC}$).
With the parameter choices of 
$(\delta_p,\delta_{\rm LOC})=(312.5,3$--$5\sigma_2 (0)), (312.5,\sigma_2 (0))$
for the concordance \LCDM and  
SCDM models,
we can identify halos with more than 30 and 58 member particles 
in each model, respectively. 
We use the constraint of $N_{core}=10$ to avoid picking up 
accidentally clustered unphysical peaks.
There is another parameter $N_{level}$, the number of
density levels which split particles of a local group into at least
$N_{level}$ particle sets.
It is desirable for each particle set
to be as small as possible to reduce underestimation of the
halo potential (see Eq. \ref{pot}).
We have found that 
there is no significant difference in the number of halo member particles 
if $N_{level}\ge 4$.
A parameter dependence study shows 
that the halo mass function obtained by the \hfind method 
depends sensitively only on $\delta_p$ at the low mass end.
Other parameters such as $\delta_{\rm LOC}$, $N_{level}$, and $N_{core}$,
are rather methodological parameters
and basically do not affect the halo finding results.

%

The main drawback of the \hfind algorithm is its memory requirement
for the fine mesh.
We need two fine meshes;
one is for the density map and the other is for the linked list of particles.
In 32-bit machines, the maximum allowable number of fine cells
is about $750^3$ which need 3.4 GB memory budget.
During the halo finding in the 1-Giga particle simulations,
we have found that a few local-cell groups require more memory than 
the 32-bit limitation.
In principle, there is no memory limitation problem on 64-bit machines
as far as there is enough physical memory allocated to each CPU.
The efficiency of the \hfind algorithm
decreases when the number of halos increases in a local particle group.
The increasing number of halos induces the computational overload
to measure the tidal radius with respect to all other massive halos.
Therefore, the measurement of a tidal radius scales as $O(N^2_h)$.
We find that about 30\% of the CPU time in the halo finding is spent 
in this job.
Algorithms faster than the current version of the \hfind algorithm
are desirable for the next largest simulations.

We apply our method to the 1-Giga particle simulations 
of the \LCDM and SCDM models. 
About four hours 
are taken for the halo finding in the simulations
of the \LCDM and SCDM models at $z=0$ on 32 IBM POWER-4 chips.
We expect the \hfind method
easily scalable to $2048^3$ or even more massive N-body simulations
due to its algorithm of dividing whole simulation particles
into multiple local particle groups.

\acknowledgments
We thank Dr. Jounghun Lee for her useful comments.
This work was supported by the Astrophysical Research Center for the Structure 
and Evolution of the Cosmos (ARCSEC) of the Korean Science and
Engineering Foundation (KOSEF) through Science Research Center (SRC) program, and
by the Basic Research Program of the KOSEF (grant no. 1999-2-113-001-5).
JK is supported in part by the Project 2003-220-00 
of Korea Astronomy Observatory.
The authors would like to acknowledge the support from 
KISTI (Korea Institute of Science and Technology Information) under 
`Grand Challenge Support Program',
and thank Dr. Sangmin Lee for the technical support.
The use of the computing system of the Supercomputing Center at KISTI
is also greatly appreciated.

\end{document}